\documentclass[aps,twocolumn,floatfix,superscriptaddress]{revtex4}
\usepackage{graphicx,times,amsmath,bm,bbm}
\begin{document}
\title{Intensity correlations, entanglement properties and ghost
imaging in multimode thermal-seeded parametric downconversion: Theory}
\author{Ivo P. Degiovanni}
\affiliation{Istituto Nazionale di Ricerca Metrologica, Torino,
Italy}
\author{Maria Bondani}
\affiliation{National Laboratory for Ultrafast and Ultraintense
Optical Science - C.N.R.-I.N.F.M., Como, Italy}
\author{Emiliano Puddu}
\affiliation{ Dipartimento di Fisica e Matematica, Universit\`a
degli Studi dell'Insubria, Como, Italia}
\affiliation{Consiglio Nazionale delle Ricerche, Istituto Nazionale per la Fisica
della Materia (C.N.R.-I.N.F.M.), Como, Italia}
\author{Alessandra Andreoni}
\affiliation{ Dipartimento di Fisica e Matematica, Universit\`a
degli Studi dell'Insubria, Como, Italia}
\affiliation{Consiglio Nazionale delle Ricerche, Istituto Nazionale per la Fisica
della Materia (C.N.R.-I.N.F.M.), Como, Italia}
\author{Matteo G. A. Paris}
\affiliation{ Dipartimento di Fisica, Universit\`a
degli Studi di Milano, I-20133 Milano, Italia}
\affiliation{I.S.I. Foundation, I-10133 Torino, Italia}
\date{\today}
\begin{abstract}
We address parametric-downconversion seeded by multimode
pseudo-thermal fields. We show that this process may be used to
generate multimode pairwise correlated states with entanglement
properties that can be tuned by controlling the seed intensities.
Multimode pseudo-thermal fields seeded parametric-downconversion
represents a novel source of correlated states, which allows one to
explore the classical-quantum transition in pairwise correlations
and to realize ghost imaging and ghost diffraction in regimes not
yet explored by experiments.
\end{abstract}
\pacs{03.67.Mn, 42.50.Dv, 42.65.Lm} \maketitle
\section{Introduction}
Ghost imaging \cite{Pit95} and ghost diffraction \cite{SSK94}
consist in the retrieval of an object transmittance pattern or its
Fourier transform, respectively by evaluating a fourth-order
correlation function at the detection planes between the field that
never interacted with the object and a correlated one transmitted by
the object. A general ghost-imaging/diffraction scheme
involves a source of correlated bipartite fields and two propagation
arms usually called test (T) and reference (R). In the T-arm, where
the object is placed, a bucket (or a point-like) detector measures
the total light transmitted by it. The R-arm contains an optical
setup suitable for reconstructing the image of the object or its
Fourier transform and a position-sensitive detector \cite{GIRS}.
\par
The correlations needed for ghost imaging and ghost diffraction may
be either quantum, as those shown by entangled states produced by
spontaneous parametric downconversion (PDC) \cite{Pit95} or
classical, as those present in the fields at the output of a
beam-splitter fed with a pair of multi-mode pseudo-thermal beams
\cite{Val05,Zha05, Fer05}. In recent years several authors discussed
analogies and differences between the two cases in terms of the
achievable visibility and of the optical configurations needed for
image reconstruction. An history of this debate from different point
of views may be found in Refs. \cite{GIRS} and references therein.
Recently, it has been suggested that the entangled nature of the
light source \cite{Dez05,Cai05,Cai05_2} may be necessary to satisfy
the ``back-propagating'' thin-lens equation, which, indeed, is
fulfilled by PDC-based ghost imaging systems. Among other things, we
prove that this claim is incorrect.
\par
In this paper, we discuss the use of a novel, PDC-based, light
source for ghost-imaging/diffraction. In our scheme (see
Fig.~\ref{expSCHEME}), the nonlinear crystal realizing PDC is seeded
by two multi-mode thermal (MMT) beams. We show that the entanglement
properties and the amount of correlation at the output may be tuned
by changing the intensities of the seeds, thus leading to a source
that can be used to investigate the transition from the classical to
the quantum regime. Besides, our novel source allows ghost-image
reconstruction with the same optical scheme used for ghost imaging
based on spontaneous PDC, with the ``back-propagating'' thin-lens
equation that is satisfied irrespective of the entanglement of the
state. We notice that the effectiveness of the setup discussed here
has been already demonstrated in the case of a crystal seeded with a
single MMT beam \cite{OPTLETT}.
\par
The paper is structured as follows. In section \ref{s:MMTPDC} we
calculate the state obtained from our PDC source with the injection
of MMT seeds on both T- and R- arms, thus revealing that the output
field on each arm maintains the statistics of the seed. In section
\ref{s:ENTCORR} we analyze both the intensity correlations between
the output beams and the entanglement properties of the overall
state. We explicitly evaluate separability thresholds in terms of
the seed intensities, and show that the condition for the existence
of nonclassical correlations in intensity measurements subsumes the
condition for inseparability, {\em i.e} sub-shot-noise correlations
are a sufficient condition for entanglement in our system. We also
show that entanglement properties of the output field are not
affected by losses taking place after the PDC interaction. In
section \ref{s:ghost} we show that the state generated in our scheme
satisfies the ``back-propagating'' thin-lens equation independently
on the seed intensities, {\em i.e.} independently on being entangled
or not, and it is suitable for realizing ghost-imaging and ghost
diffraction experiments. Finally, Section \ref{s:out} closes the
paper with some concluding remarks.
\section{Parametric down-conversion with thermal seeds}
\label{s:MMTPDC}
\begin{figure}[h]
\includegraphics[width=0.45\textwidth, angle=0]{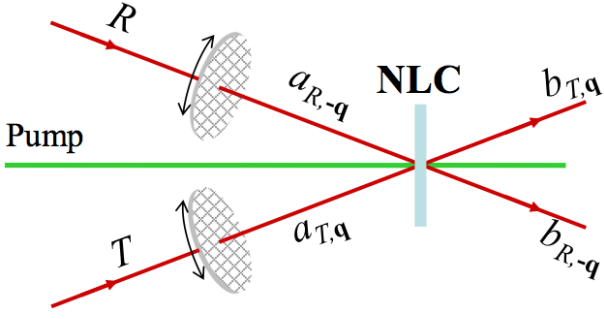}
\caption{Schematic diagram of the nonlinear interaction. $T$ and $R$
are the Test and Reference arms of the setup.}\label{expSCHEME}
\end{figure}

The interaction scheme we are going to consider is schematically
depicted in Fig.~\ref{expSCHEME}. It consists in a nonlinear
$\chi^{(2)}$ crystal pumped by a monochromatic non-depleted
plane-wave propagating along the $z$-axis. The Hamiltonian
describing the resulting parametric process is given by
\begin{equation} \label{hamiltonian}
H_{I}=\int\!\! \mathrm{d}^{2}\mathbf{x}\int_{0}^{L}
\!\!\!\mathrm{d}z\,
\chi^{(2)}
 E_{p}(\mathbf{x}, t){a}_{T}(\mathbf{x},
t){a}_{R}(\mathbf{x}, t) + h.c.
\end{equation}
$L$ being the crystal length and $\chi^{(2)}$ the nonlinear
susceptibility. The pump-field may be written as
$E_{p}(\mathbf{x},
t)=\mathcal{E}_{p}\exp\left[i(\Omega_{p} t -K_{p}z )\right]$
\cite{mandel_wolf}.
\par
We can write the interacting quantum fields as
\begin{equation}
{a}_{j}(\mathbf{x}, t)\propto \sum_{\mathbf{q}_{j},
\nu_{j}}{a}_{j,\mathbf{q}_{j}, \nu_{j}} e^{i[K_{j,z}z +
\mathbf{q}_{j}\cdot\mathbf{ r}- (\Omega_{j}+\nu_{j})t]} ~~~~ (j=R,T)
\end{equation}
where: $K_{j,z}=\sqrt{K_{j}^{2}-\mathbf{q}_{j}^{2}}$,
$\mathbf{q}_{j}$ being the transverse momentum, $K_{j}= n_{j}
(\Omega_{j}+\nu_{j})/c$, $n_{j}$ the index of refraction,
$\Omega_{j}$ the selected central frequency in channel $j$,
$\nu_{j}$ the frequency displacement with respect to $\Omega_{j}$,
and $c$ is the speed of light in the vacuum. The commutation
relation of the quantum fields are
\begin{eqnarray}\label{commrel}
[{a}_{j,\mathbf{q}, \nu },{a}_{j',\mathbf{q}',
\nu' }^{\dag}]&=&\delta_{j,j'}
\delta_{\mathbf{q},\mathbf{q}'}\delta_{\nu,\nu'} ~~~~(j,j'=R,T)
\\ \nonumber [{a}_{j,\mathbf{q}, \nu },{a}_{j',\mathbf{q}', \nu' }]&=&0
\end{eqnarray}
The evolution of a quantum system induced by the interaction
Hamiltonian (\ref{hamiltonian}) is described by the unitary
operator ${U}=\exp(-i \hbar^{-1}\int\!{H}_{I}\,\mathrm{d}t)$, where
\begin{align}
\label{operatorU}
 -\frac{i}{\hbar} \int \! {H}_{I}\, \mathrm{d}t
 = i \sum_{\mathbf{q}, \nu} \kappa_{\mathbf{q},
\nu} {a}_{T, \mathbf{q}, \nu }{a}_{R, -\mathbf{q},
-\nu } + h.c.
\end{align}
where $\kappa_{\mathbf{q}, \nu} \propto \mathrm{sinc}\left[(K_{p}-
K_{T,z}-K_{R,z})L/2 \right]$. To obtain Eq. (\ref{operatorU}) we
have exploited the conservation of energy at the central
wavelength $\Omega_{p}=\Omega_{T}+\Omega_{R}$ obtaining
$\nu_{T}=-\nu_{R}=\nu$, and the conservation of transverse
momentum $\mathbf{q}_{T}=-\mathbf{q}_{R}=\mathbf{q}$.
As, according to Eq.~(\ref{operatorU}), the extension to the
non-monochromatic case is, in most of cases, straightforward, in
the following analysis we will focus on the monochromatic emission
at the frequencies $\Omega_{R}$ and $\Omega_{T}$ and
hence we will drop the subscript $\nu$ from the variables.
\par\noindent
The operator ${U}$ can be rewritten in terms of the operators
${S}_{\mathbf{q}}=(\kappa_{\mathbf{q}} {a}_{T,\mathbf{q}} {a}_{R,
\mathbf{-q}} + h.c.)$ as ${U}= \exp \left(i \sum_{\mathbf{q}}
{S}_{\mathbf{q}} \right)$. According to the commutation relations
(\ref{commrel}), we have $[{S}_{\mathbf{q}},{S}_{\mathbf{q'}}]=0$,
and therefore ${U}=\bigotimes_{\mathbf{q}}e^{i {S}_{\mathbf{q}}}$,
{\em i.e.} the interaction establishes pairwise correlations among
the modes.
\par
In our analysis we focus on the case in which both
the T- and R-arms are seeded with MMT beams
\begin{align}\label{rhoin}
{\rho}_{in}&=\bigotimes_{\mathbf{q}}
{\rho}_{T,\mathbf{q}}\otimes {\rho}_{R,-\mathbf{q}}\\
{\rho}_{j,\mathbf{q}}&=\sum_{n=0}^{\infty}P_{j,\mathbf{q}}(n)~
|n\rangle_{j,\mathbf{q}}{}_{j,\mathbf{q}}\langle n |\nonumber\:,
\end{align}
where $j=R,T$ and $\left| n \right\rangle_{j,\mathbf{q}}$
denotes the Fock number basis for the mode $\mathbf{q}$
of the $j$-arm. The thermal profile of the input is given by
$$P_{j,\mathbf{q}}(n)=\mu_{j,\mathbf{q}}^{n}(1+\mu_{j,
\mathbf{q}})^{-n-1}\,,$$
$\mu_{j,\mathbf{q}}$ being the average photon number per mode.
The density matrix at the output is given by
\begin{equation}\label{rhooutq}
{\rho}_{out} =
{U}{\rho}_{in}{U}^{\dagger}=
\bigotimes_{\mathbf{q}} e^{i {S}_{\mathbf{q}}}\:
{\rho}_{T,\mathbf{q}} \otimes {\rho}_{R,\mathbf{-q}}\:
e^{-i {S}_{\mathbf{q}}}\:,
\end{equation}
\par
According to \cite{traux}, it is possible to ``disentangle'' $e^{i
{S}_{\mathbf{q}}}$ by using the two-boson representation of the
SU(1,1) algebra as
\begin{align} \label{su11}
e^{i {S}_{\mathbf{q}}} =
&\exp\left\{
{\zeta_{\mathbf{q}}{a}_{T,\mathbf{q}}^{\dagger} {a}_{R,\mathbf{-q}}^{\dagger} }
\right\}
\nonumber \\ \times &
\exp\left\{
{ -\eta_{\mathbf{q}}({a}_{T,\mathbf{q}}^{\dagger}{a}_{T,\mathbf{q}}+
{a}_{R,\mathbf{-q}}^{\dagger}{a}_{R,\mathbf{-q}}
+ 1) }
\right\}
\nonumber \\ \times &
\exp\left\{
{-\zeta_{\mathbf{q}}^{*} {a}_{T,\mathbf{q}} {a}_{R,\mathbf{-q}}}
\right\}
\end{align}
where $\zeta_{\mathbf{q}}=-i e^{-i \varphi_{\mathbf{q}}}
\tanh(|\kappa_{\mathbf{q}}|)$, $\eta_{\mathbf{q}}=\ln
[\cosh|\kappa_{\mathbf{q}}|]$, and $e^{i
\varphi_{\mathbf{q}}}=\kappa_{\mathbf{q}}/|\kappa_{\mathbf{q}}|$.
\begin{widetext}
Equation (\ref{su11}) implies that
\begin{align} \label{Umn}
e^{i {S}_{\mathbf{q}}}|n\rangle_{T,\mathbf{q}}\otimes|m\rangle_{R,-\mathbf{q}}
= \sum_{k=0}^{\min\{m,n\}} \sum_{l=0}^{\infty}C_{\mathbf{q}}(m,n,k,l)\:
|n-k+l\rangle_{T,\mathbf{q}}\otimes|m-k+l\rangle_{R,-\mathbf{q}}
\end{align}
with
\begin{align}
C_{\mathbf{q}}(m,n,k,l)= e^{ -\eta_{\mathbf{q}}(n+m-2 k + 1) }
\frac{\sqrt{n! m! (n-k+l)!(m-k+l)!}}{k! l! (n-k)! (m-k)!}
\zeta_{\mathbf{q}}^{l}(-\zeta_{\mathbf{q}}^{*})^{k}
\end{align}
By substituting Eq. (\ref{Umn}) in Eq. (\ref{rhooutq}), we obtain the output state
\begin{align} \label{rhooutfinal}
{\rho}_{out}=&  \bigotimes_{\mathbf{q}}
\sum_{nm} P_{T,\mathbf{q}}(n) ~ P_{R,-\mathbf{q}}(m)
\sum_{k_{1},k_{2}=0}^{\min\{m,n\}} \sum_{l_{1}, l_{2}=0}^{\infty}
 C_{\mathbf{q}}(m,n,k_{1},l_{1})~C_{\mathbf{q}}(m,n,k_{2},l_{2})^{*}
\times \nonumber \\
&\quad |n-k_{1}+l_{1}\rangle_{T,\mathbf{q}}
{}_{T,\mathbf{q}}
\langle n-k_{2}+l_{2}|\otimes|m-k_{1}+l_{1}\rangle_{R,-\mathbf{q}}{}_{R,-\mathbf{q}}
\langle m-k_{2}+l_{2}|
\end{align}
As expected, the first moments of the photon distribution for each
mode are those of a thermal statistics
\begin{eqnarray} \label{averages}
\langle {n}_{T,\mathbf{q}} \rangle &=&
\mathrm{Tr}({\rho}_{out}
{a}_{T,\mathbf{q}}^{\dagger}{a}_{T,\mathbf{q}})=\mu_{T,\mathbf{q}}+
n_{\mathrm{PDC},\mathbf{q}}(1+ \mu_{T,\mathbf{q}}+
\mu_{R,-\mathbf{q}})
\nonumber\\
\langle {n}_{R,-\mathbf{q}} \rangle &=&
\mathrm{Tr}({\rho}_{out}
{a}_{R,-\mathbf{q}}^{\dagger}{a}_{R,-\mathbf{q}})=\mu_{R,-\mathbf{q}}+
n_{\mathrm{PDC},\mathbf{q}}(1+ \mu_{T,\mathbf{q}}+
\mu_{R,-\mathbf{q}})
\\
\langle (\Delta n_{T,\mathbf{q}})^{2}\rangle &=& \langle
{n}_{T,\mathbf{q}} \rangle (\langle {n}_{T,\mathbf{q}} \rangle+1)
 \nonumber\\
\langle (\Delta{n}_{R,\mathbf{-q}})^{2}\rangle &=& \langle
{n}_{R,-\mathbf{q}} \rangle (\langle {n}_{R,-\mathbf{q}} \rangle +
1) \nonumber
\end{eqnarray}
where $\langle {O} \rangle=
\mathrm{Tr}[{O}{\rho}_{out}]=\mathrm{Tr}[{U}^{\dag}{O}{U}{\rho}_{in}]$,
$\Delta{O}= {O}- \langle {O} \rangle $ and
$n_{\mathrm{PDC},\mathbf{q}}=\sinh^2|\kappa_{\mathbf{q}}|$ is the
average number of photons due to spontaneous PDC.
\end{widetext}
Notice that the case of vacuum inputs, ${\rho}_{in}=
|0\rangle\langle 0|_{T}\otimes|0\rangle\langle 0|_{R}$,
corresponds to spontaneous downconversion, {\em i.e.} to the
generation of twin-beam, whereas the case of a single MMT on
one arm and the vacuum on the other,
${\rho}_{in}=\bigotimes_{\mathbf{q}}
\left(|0\rangle\langle 0|_{T,\mathbf{q}}\otimes
{\rho}_{R,-\mathbf{q}} \right)$ corresponds to the state
considered in Ref. \cite{OPTLETT}.
\par
In the Heisenberg description, the modes
after the interaction with the crystal are given by
${b}_{j,\mathbf{q}}={U}^{\dag}{a}_{j,\mathbf{q}}{U}$, {\em i.e}
\begin{equation}
{b}_{j,\mathbf{q}}=\mathcal{U}_{\mathbf{q}}{a}_{j,\mathbf{q}}+ e^{i
\varphi_{\mathbf{q}}}\mathcal{V}_{\mathbf{q}}{a}_{j',\mathbf{-q}}^{\dag}
~~ (j,j'=R,T , ~j \neq j') \label{aout}
\end{equation}
where $\mathcal{U}_{\mathbf{q}}=\cosh|\kappa_{\mathbf{q}}|$ and
$\mathcal{V}_{\mathbf{q}}= \sinh|\kappa_{\mathbf{q}}|$ (and
obviously $\mathcal{U}_{\mathbf{q}}=\mathcal{U}_{\mathbf{-q}}$,
$\mathcal{V}_{\mathbf{q}}=\mathcal{V}_{\mathbf{-q}}$, and
$\varphi_{\mathbf{q}}=\varphi_{\mathbf{-q}}$).
\section{Entanglement and intensity correlations}
\label{s:ENTCORR}
In this section we address intensity correlations and entanglement
properties of the beams generated in our scheme. As we will see,
the amount of nonclassical correlations and entanglement may be tuned
upon changing the intensity of the thermal seeds and there exist
thresholds for the appearance of those nonclassical features. On the
other hand, the index of total correlations (either classical or quantum)
is a monotonically increasing function of the both the seed and the
PDC energy.
\subsection{Entanglement and separability}
The downconversion process is known to provide pairwise entanglement
between signal and idler beams. In our notations the (possibly) entangled
mode are $a_{T,\mathbf{q}}$ and $a_{R,\mathbf{-q}}$. In the spontaneous
process the output state is entangled for any value of the parametric
gain ({\em i.e} for any value of the crystal susceptibility, length $\dots$)
whereas in the case of a thermally seeded crystal the degree of entanglement
crucially depends on the intensity of the seeds.
\par
Since thermal states are Gaussian and the PDC Hamiltonian is
bilinear in the field modes, the overall output state is also
Gaussian. Therefore the entanglement properties may be evaluated by
checking the positivity of the partial transpose (PPT condition),
which represents a sufficient and necessary condition for
separability for Gaussian pairwise mode entanglement \cite{simon00}.
Gaussian states are completely characterized by their covariance
matrix. In this context let us introduce the ``position''(-like)
operators ${X}$ and ``momentum''(-like) operators ${Y}$
\begin{eqnarray}
{X}_{j,\mathbf{q}}&=&\frac{{a}_{j,\mathbf{q}}+{a}_{j,\mathbf{q}}^{\dag}}{\sqrt{2}}
 \nonumber\\
{Y}_{j,\mathbf{q}}&=&\frac{{a}_{j,\mathbf{q}}-{a}_{j,\mathbf{q}}^{\dag}}{i\sqrt{2}}
~~~ (j=R,T )\label{XY}
\end{eqnarray}
Introducing the vector operator
\begin{eqnarray}\mathbf{{\xi}}=({X}_{T,\mathbf{q}_{1}},{Y}_{T,\mathbf{q}_{1}},
{X}_{R,\mathbf{-q}_{1}},{Y}_{R,\mathbf{-q}_{1}},...)^{T}
\end{eqnarray}
with $m=1,2,...,\infty$, from the commutation relations
(\ref{commrel}) gives
\begin{equation} \label{commrel2}
\left[{\xi}_{\alpha} ,  {\xi}_{\beta} \right]= i
\Omega_{\alpha,\beta}
\end{equation}
where $\mathbf{\Omega}= \bigoplus_m \mathbf{\omega} \oplus \mathbf{\omega}$
and $\mathbf{\omega}$ is the symplectic matrix
\begin{equation} \label{symplectic}
\mathbf{\omega} = \left(
\begin{array}{cc}
0 & 1 \\
-1 & 0
\end{array} \right)
\end{equation}
\par
The covariance matrix $\mathbf{\mathrm{V}}$ is calculated as
$\mathrm{V}_{\alpha,\beta}=2^{-1} \langle \{
\Delta{\xi}_{\alpha},\Delta{\xi}_{\alpha} \}\rangle$,
where $\{ {O}_{1},{O}_{2} \}$ denotes the
anti-commutator. Uncertainty relation among the position and momentum
operators impose a constraint on the covariance matrix,
$\mathbf{V}+\frac{1}{2}\mathbf{\Omega} \geq 0$,
corresponding to the positivity of the state.
The input-output relations for position and momentum operators are
calculated according to Eq.s (\ref{aout}), obtaining
\begin{align}
{U}^{\dag}{X}_{j,\mathbf{q}}{U}  &=
\mathcal{U}_{\mathbf{q}}{X}_{j,\mathbf{q}}+
\mathcal{V}_{\mathbf{q}}
\label{XYout}
{X}_{j',\mathbf{-q}}  \\
{U}^{\dag} {Y}_{j,\mathbf{q}}{U} &=\mathcal{U}_{\mathbf{q}}{Y}_{j,\mathbf{q}}-
\mathcal{V}_{\mathbf{q}}{Y}_{j',\mathbf{-q}} \:\:(j,j'=R,T ,
~ j \neq j')\,.
\nonumber
\end{align}
\par
Without any loss of generality, in the derivation of Eq.s
(\ref{XYout}) we set $\varphi_{\mathbf{q}}=0$ which, in turn,
corresponds to a proper choice of the phase, or, equivalently to a
proper redefinition of the operators ${a}_{j,\mathbf{q}}$
corresponding to a rotation of the phase space.
From Eq.s (\ref{XYout}) we calculate the covariance matrix
\begin{equation} \label{Vcov}
\mathbf{V} =\bigoplus_{m=1}^{\infty} \mathbf{V}_{\mathbf{q}_{m}}=
\left(
\begin{array}{cccc}
\mathbf{V}_{\mathbf{q}_{1}} & \mathbf{0} & \mathbf{0} & ... \\
\mathbf{0}  & \mathbf{V}_{\mathbf{q}_{2}} & \mathbf{0}  & ...   \\
\mathbf{0} & \mathbf{0}   & \mathbf{V}_{\mathbf{q}_{3}} & ...   \\
\vdots  & \vdots  & \vdots  & \ddots
\end{array} \right)
\end{equation}
with
\begin{equation} \label{Vcal}
\mathbf{V}_{\mathbf{q}}= \left(
\begin{array}{cccc}
\mathcal{A}_{\mathbf{q}} & 0 & \mathcal{C}_{\mathbf{q}} & 0 \\
0  & \mathcal{A}_{\mathbf{q}} & 0  & -\mathcal{C}_{\mathbf{q}}  \\
\mathcal{C}_{\mathbf{q}} & 0   & \mathcal{B}_{\mathbf{q}} & 0   \\
0  & -\mathcal{C}_{\mathbf{q}}  & 0 & \mathcal{B}_{\mathbf{q}}
\end{array} \right)
\end{equation}
where
\begin{eqnarray}
\mathcal{A}_{\mathbf{q}}&=& [\mathcal{U}_{ \mathbf{q}}^{2}(2
\mu_{T,\mathbf{q}}+1)+ \mathcal{V}_{\mathbf{q}}^{2}(2
\mu_{R,-\mathbf{q}}+1)]/2, \nonumber
\\ \mathcal{B}_{\mathbf{q}}&=&
[\mathcal{U}_{\mathbf{q}}^{2}(2 \mu_{R,-\mathbf{q}}+1)+
\mathcal{V}_{\mathbf{\mathbf{q}}}^{2}(2 \mu_{T,\mathbf{q}}+1)]/2, \nonumber
\\ \mathcal{C}_{\mathbf{q}}&=& \mathcal{U}_{\mathbf{q}}
\mathcal{V}_{\mathbf{q}} (\mu_{T,\mathbf{q}}+\mu_{R,-\mathbf{q}}+1).
\label{primati}
\end{eqnarray}
$\mathbf{V}$ satisfies the uncertainty relations
ensuring the positivity of ${\rho}_{out}$.
\par
In order to check whether and when the state ${\rho}_{out}$
is entangled we apply the PPT criteria for Gaussian entanglement
\cite{simon00}. For instance, we apply the positive map
$\mathcal{L}_{R,\mathbf{-q'}}$ to the state ${\rho}_{out}$.
$\mathcal{L}_{R,\mathbf{-q'}}({\rho}_{out})$ is the
transposition (complex conjugation) only of the subspace
$\mathcal{H}_{R,\mathbf{-q'}}$ corresponding to the mode
$R,\mathbf{-q'}$. Simon showed that this corresponds to calculate
the covariance matrix $\widetilde{\mathbf{V}}$, where all the
matrix-blocks $\mathbf{V}_{\mathbf{q}}$ remain the same excepts the
matrix $\mathbf{V}_{\mathbf{q'}}\rightarrow
\widetilde{\mathbf{V}}_{\mathbf{q'}} $.
$\widetilde{\mathbf{V}}_{\mathbf{q'}}$ is calculated with a sign
change in the $R,\mathbf{-q'}$ momentum variable
(${Y}_{R,\mathbf{-q'}}\rightarrow
-{Y}_{R,\mathbf{-q'}} $), while the other momentum and
position variables remain unchanged
(${X}_{T,\mathbf{q'}}\rightarrow {X}_{T,\mathbf{q'}}
$, ${Y}_{T,\mathbf{q'}}\rightarrow
{Y}_{T,\mathbf{q'}} $, and
${X}_{R,\mathbf{-q'}}\rightarrow
{X}_{R,\mathbf{-q'}} $). Thus we obtain
\begin{align}
\label{Vcaltilde}
\widetilde{\mathbf{V}}_{\mathbf{q'}}= \left(
\begin{array}{cccc}
\mathcal{A}_{\mathbf{q'}} & 0 & \mathcal{C}_{\mathbf{q'}} & 0 \\
0  & \mathcal{A}_{\mathbf{q'}} & 0  & \mathcal{C}_{\mathbf{q'}}  \\
\mathcal{C}_{\mathbf{q'}} & 0   & \mathcal{B}_{\mathbf{q'}} & 0   \\
0  & \mathcal{C}_{\mathbf{q'}}  & 0 & \mathcal{B}_{\mathbf{q'}}
\end{array} \right),
\end{align}
where $\mathcal{A}_{\mathbf{q'}}$,  $\mathcal{B}_{\mathbf{q'}}$ and
$\mathcal{C}_{\mathbf{q'}}$ are defined in Eq.s (\ref{primati}).
According to PPT criteria, the separability of
${\rho}_{out}$, is guaranteed by the positivity of
$\mathcal{L}_{R,-q'}({\rho}_{out})$, i.e.
\begin{equation} \label{PPT}
\widetilde{\mathbf{V}}+\frac{1}{2}\mathbf{\Omega} \geq 0\:.
\end{equation}
Inequality (\ref{PPT})
corresponds to
\begin{equation} \label{autovalore}
\mu_{T,\mathbf{q}'} \mu_{R,-\mathbf{q}'}- n_{\mathrm{PDC},\mathbf{q}'} (1+\mu_{T,\mathbf{q}'} +
\mu_{R,-\mathbf{q}'})\geq 0.
\end{equation}
We observe that the spontaneous PDC corresponds to the situation
with $\mu_{T,\mathbf{q}'}=\mu_{R,-\mathbf{q}'}=0$, thus
${\rho}_{out}$ is entangled. Also the case considered in
Ref. \cite{OPTLETT}, a MMT seeded PDC only on one arm (i.e.,
$\mu_{R,-\mathbf{q}'}=0$) is always entangled. On the contrary, in the
case of MMT seeded PDC on both arms, the inequality
(\ref{autovalore}) introduces a threshold. For instance, if we consider a MMT seed
with the same mean number of photon per mode, $\mu$, only when the
inequality $\mu^{2}\geq  n_{\mathrm{PDC},\mathbf{q}}(1 + 2
\mu)$ is satisfied, ${\rho}_{out}$ is separable. It is
noteworthy to observe that if the PPT is applied to any other
subspaces the inequality obtained are analogous to Eq.
(\ref{autovalore}), and thus the result is the same.
\subsection{Separability and losses}
Her we address the problem of the effect of the losses
on the separability of the state (\ref{rhooutfinal}). In fact the
presence of losses, e.g. internal reflection or absorption in the
nonlinear crystal, may modify the quantum properties of the state,
in particular the transition from entanglement to separability (in
the absence of losses given by the Eq.~(\ref{autovalore})).
\par
Losses in a quantum channel can be modeled by a beam splitter in one
port of which the quantum channel is injected while the vacuum
enters the other port. The model implies that Gaussian states after
interaction are still Gaussian states due to the bi-linearity of the
beam-splitter Hamiltonian. Thus also in the presence of losses, the
covariance matrix completely describes the quantum state. If we
consider an overall transmission factor $\tau$ on both channels we
obtain the covariance matrix $\mathbf{V}_{\tau}= \tau \mathbf{V}+
(1-\tau) \mathbf{1}/2 $. The form of the covariance matrix
$\mathbf{V}_{\tau}$ is completely analogous to Eq. (\ref{Vcov}),
where the block matrices $\mathbf{V}_{\mathbf{q}}$ are substituted
with the block matrices $\mathbf{V}_{\tau,\mathbf{q}}$.
$\mathbf{V}_{\tau,\mathbf{q}}$ has the same structure of
$\mathbf{V}_{\mathbf{q}}$ in Eq. (\ref{Vcal}), where
$\mathcal{A}_{\mathbf{q}}$, $\mathcal{B}_{\mathbf{q}}$, and
$\mathcal{C}_{\mathbf{q}}$ are substituted by
$\mathcal{A}_{\tau,\mathbf{q}}= \{1+ 2 ~ \tau~ [\mathcal{U}_{
\mathbf{q}}^{2}~ \mu_{T,\mathbf{q}}+ \mathcal{V}_{\mathbf{q}}^{2}(
\mu_{R,-\mathbf{q}}+1)]\}/2 $, $\mathcal{B}_{\tau,\mathbf{q}}= \{1+
2 ~ \tau~ [\mathcal{U}_{ \mathbf{q}}^{2}~ \mu_{R,-\mathbf{q}}+
\mathcal{V}_{\mathbf{q}}^{2}( \mu_{T,\mathbf{q}}+1)]\}/2 $, and
$\mathcal{C}_{\tau,\mathbf{q}}= \tau ~\mathcal{C}_{\mathbf{q}} $,
respectively. Thus, following the same line of thought of the
previous section we obtain the covariance matrix
$\tilde{\mathbf{V}}_{\tau}$, corresponding to the partial
transposition of the state. According to PPT
separability criteria, the state is separable if and only if the
inequality $\widetilde{\mathbf{V}}_{\tau}+\frac{1}{2}\mathbf{\Omega}
\geq 0$ is fulfilled. This condition can be rewritten as
\begin{equation} \label{autovalore2}
\tau^{2}[\mu_{T,\mathbf{q}'}
\mu_{R,-\mathbf{q}'}- n_{\mathrm{PDC},\mathbf{q}'} (1+\mu_{T,\mathbf{q}'} +
\mu_{R,-\mathbf{q}'})]\geq 0.
\end{equation}
Since Ineq.~(\ref{autovalore2}) is fully equivalent to
Ineq.~(\ref{autovalore}), we conclude that losses do not affect the
entanglement properties of the state in Eq.
(\ref{rhooutfinal}).
\subsection{Intensity correlations}
We now evaluate the pairwise intensity correlations owned by
the generated beams. In addition, we analyze the connections between
threshold for separability and the threshold required to have nonclassical
correlations.  As we will see a state obtained by thermally seeded PDC
that exhibits sub shot-noise correlations is entangled, whereas the
converse is not necessarily true. In other words, the existence of
nonclassical intensity correlations is a sufficient condition for
entanglement.
\par
\begin{figure}[h]
\includegraphics[width=0.45\textwidth,angle=0]{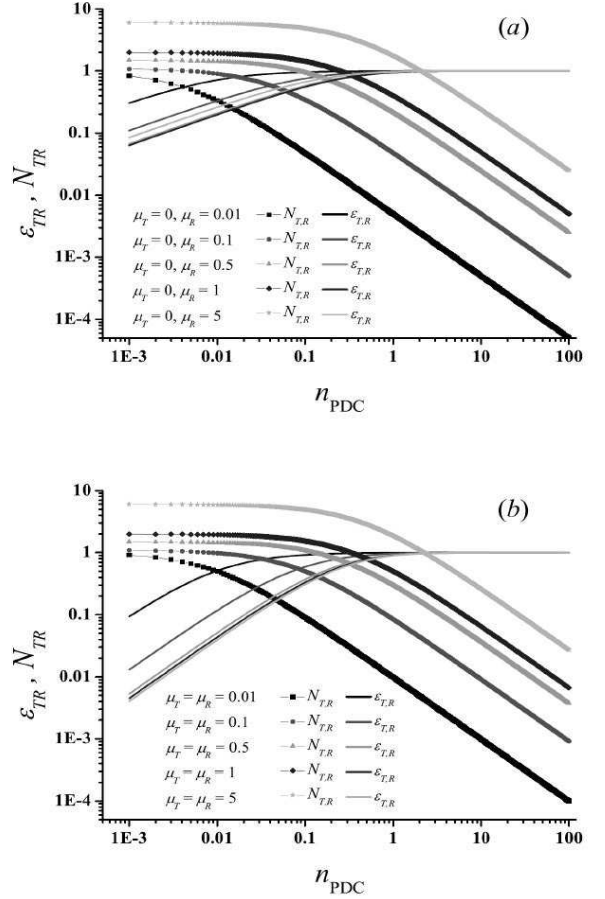}
\caption{The index of total correlations $\gamma_{T,R}$ (solid line)
and the noise reduction factor $NRF_{T,R}$ (line
plus symbol) as a function of $n_\mathrm{PDC}$ for the cases of
($a$): $\mu_{T}=0$ and $\mu_{R}\neq 0$ and ($b$)
$\mu_{T}=\mu_{R}\neq 0$. The values chosen for the parameters are
indicated in the figures.}\label{eps}
\end{figure}

The normalized index of intensity correlation between a pair of modes
$a_{j,\mathbf{q}}$ and $a_{j',\mathbf{q'}}$ is defined as
\begin{eqnarray}
\gamma_{j,j'}(\mathbf{q},\mathbf{q'}) =
\frac{\Gamma_{j,j'}(\mathbf{q},\mathbf{q'})}{\sqrt{\langle (\Delta
n_{T,\mathbf{q}})^{2}\rangle \langle
(\Delta{n}_{R,\mathbf{-q}})^{2}\rangle }} \label{epsPH}\; .
\end{eqnarray}
where the correlation term is given by
\begin{eqnarray}
\Gamma_{j,j'}(\mathbf{q},\mathbf{q'}) = \left\langle \Delta
n_{j,\mathbf{q}} \Delta n_{j',\mathbf{q'}} \right\rangle
\label{diff}\;.
\end{eqnarray}
Upon evaluating the first moments as we did in Eq. (\ref{averages})
we have, for the pair of modes $a_{T,\mathbf{q}}$ and
$a_{R,\mathbf{-q}}$,
\begin{eqnarray}
\Gamma_{T,R}(\mathbf{q},-\mathbf{q}) &=& n_{\mathrm{PDC},\mathbf{q}}
\left(1+n_{\mathrm{PDC},\mathbf{q}}\right)
\left(1+\mu_{T,\mathbf{q}} +\mu_{R,-\mathbf{q}}\right)^2  \nonumber
\\ &=& \mathcal{C}_{\mathbf{q}}^2. \label{epsTR}
\end{eqnarray}
A nonzero value of $\Gamma_{T,R}$, and hence of $\gamma_{T,R}$,
indicates the presence of correlations between the considered modes.
Perfect correlations correspond to $\gamma_{T,R}=1$. Note that
$\gamma_{T,R}$ is an increasing function of $n_\mathrm{PDC}$, and
does not undergo any threshold.
In Fig~\ref{eps} we plot $\gamma_{T,R}$ (solid lines) as a function
of $n_{\mathrm{PDC}}$ in two different conditions, namely
$\mu_{T}=0$ and $\mu_{R}\neq 0$ (panel ($a$)) and
$\mu_{T}=\mu_{R}\neq 0$ (panel ($b$)). As expected $\gamma_{T,R}$
approaches unit irrespectively of the mean values of the seeding
thermal fields as soon as $n_{\mathrm{PDC}}$ becomes relevant.
For large $n_{\mathrm{PDC},\mathbf{q}}$ the index of correlation approaches unit
as follows
\begin{align}
\gamma_{T,R}(\mathbf{q},-\mathbf{q})  \simeq 1- \frac12 \frac{
\mu_{T,\mathbf{q}} +\mu_{R,-\mathbf{q}} + 2 \mu_{T,\mathbf{q}}
\mu_{R,-\mathbf{q}}}{ \left(1+\mu_{T,\mathbf{q}}
+\mu_{R,-\mathbf{q}}\right)^2} \frac1{
n^2_{\mathrm{\mathrm{PDC}},\mathbf{q}}} \:.
\end{align}
In the two cases
$\mu_{T,\mathbf{q}}=\mu\gg 1$ and $\mu_{R,-\mathbf{q}}=0$
(or viceversa) and
$\mu_{T,\mathbf{q}}=\mu_{R,-\mathbf{q}}=\mu\gg 1$
we have, respectively
\begin{align}
\gamma_{T,R}(\mathbf{q},-\mathbf{q})  & \simeq 1
- \frac{1}{(1+n_{\mathrm{PDC},\mathbf{q}})n_{\mathrm{PDC},\mathbf{q}}}\frac{1}{2\mu} \\
\gamma_{T,R}(\mathbf{q},-\mathbf{q})  & \simeq 1 - \frac{1}{(1+2
n_{\mathrm{PDC},\mathbf{q}})^2} + O\left(\frac{1}{\mu^2}\right) \nonumber\:.
\end{align}
\par
The nonclassical nature of this pairwise correlation may be assessed by the
quantity \cite{agliati}
\begin{align}
NRF_{T,R}(\mathbf{q}) = \frac{ \langle (\Delta
n_{T,\mathbf{q}})^{2}\rangle + \langle
(\Delta{n}_{R,\mathbf{-q}})^{2}\rangle -2 \Gamma_{T,R}(\mathbf{q},
-\mathbf{q})}{\langle {n}_{T,\mathbf{q}} \rangle + \langle
{n}_{R,-\mathbf{q}} \rangle } \:
\end{align}
which is usually referred to as ``the noise reduction factor''. A noise
reduction, $NRF_{T,R}(\mathbf{q})<1$, indicates the presence of
nonclassical correlations. The value $NRF_{T,R}(\mathbf{q})=1$ is
usually called ``shot-noise limit'' and corresponds to the
case of a pair of uncorrelated coherent signals.
By substituting the result for our system, we get
\begin{align}
NRF_{T,R} (\mathbf{q})=\frac{
\mu_{T,\mathbf{q}}\left(1+\mu_{T,\mathbf{q}}\right)+
\mu_{R,-\mathbf{q}}\left(1+\mu_{R,-\mathbf{q}}\right)}{
\mu_{T,\mathbf{q}} + \mu_{R,-\mathbf{q}} + 2
n_{\mathrm{PDC},\mathbf{q}} \left(1+\mu_{T,\mathbf{q}}
+\mu_{R,-\mathbf{q}}\right)}
\end{align}
We have $NRF_{T,R}(\mathbf{q})<1$ if
\begin{align}
n_{\mathrm{PDC},\mathbf{q}} > \frac12 \frac{\mu^2_{T,\mathbf{q}}
+\mu^2_{R,-\mathbf{q}}}{1+\mu_{T,\mathbf{q}} +\mu_{R,-\mathbf{q}}}
\:,
\end{align}
which subsumes the separability threshold of Eq.~(\ref{autovalore}) and
individuates the same region for $\mu_{T,\mathbf{q}} =
\mu_{R,-\mathbf{q}}$. Therefore, for thermally seeded PDC, sub-shot
noise correlations imply entanglement \cite{subshot}.
In Fig.~\ref{eps} we also plot $NRF_{T,R}$ as a function of
$n_{\mathrm{PDC}}$ for the same parameters used for $\gamma_{T,R}$.
As expected, the figure shows that $NRF_{T,R}$ crosses the
shot-noise level at different values of $n_{\mathrm{PDC}}$ that
depend on the mean values of the thermal seeds, thus confirming the
intuition that in order to achieve sub-shot noise correlations in
the presence of two thermal seeds we need to have a PDC process
strong enough.
\section{MMT-PDC based ghost imaging and ghost diffraction}
\label{s:ghost}
The bipartite state obtained by the nonlinear process described
above is suitable for applications to ghost-imaging/diffraction
protocols.
Ghost-imaging and ghost diffraction protocols rely on the capability
of retrieving an object transmittance pattern and its Fourier
transform, respectively, by the evaluation of a fourth-order
correlation function at the detection planes of a light field that
has never interacted with the object and a correlated one transmitted
by the object. We consider the schemes depicted in
Fig.~\ref{setupGHOST}. An object, described by the transmission
function $t(\mathbf{x}''_{T})$, is inserted in the T-arm on the
plane $\mathbf{x}''_{T}$ and a bucket detector measures the total
light, ${I}_T$, transmitted by the transparency. The R-arm contains
an optical setup suitable for reconstructing either the image of the
object or its Fourier transform and a position-sensitive detector that
measures the local ${I}_R(\mathbf{x}_R)$. The procedure for calculating
the correlation function between the light detected in the two arms
of the setup is equivalent to evaluating first the correlation
function between ${I}_R(\mathbf{x}_R)$ and ${I}_T(\mathbf{x}_T)$:
\begin{equation}
G^{(2)}(\mathbf{x}_R, \mathbf{x}_T) = \langle \Delta{I}_R
(\mathbf{x}_R) \Delta{I}_T (\mathbf{x}_T) \rangle   , \label{eqTH1}
\end{equation}
and then integrating over all the values of $\mathbf{x}_T$
\begin{equation}
\mathcal{G}^{(2)}(\mathbf{x}_R) = \int{d\mathbf{x}_T ~
G^{(2)}(\mathbf{x}_R,\mathbf{x}_T)}, \label{Corrfunc}
\end{equation}
where $\langle {I}_j (\mathbf{x}_i)\rangle = \langle
{c}_j^\dagger(\mathbf{x}_j) {c}_j(\mathbf{x}_j) \rangle $ ($j=R,T$)
is the mean intensity of the {\it j}-th beam at the detection plane,
with hereinafter $\langle ... \rangle = \mathrm{Tr}\left(...
\rho_{in} \right)$, and $\langle {I}_R (\mathbf{x}_R) {I}_T
(\mathbf{x}_T) \rangle = \langle {c}_R^\dagger (\mathbf{x}_R) {c}_R
(\mathbf{x}_R) {c}_T^\dagger (\mathbf{x}_T) {c}_T (\mathbf{x}_T)
\rangle$.
\par
The connection between the field operators at the detection planes
and those at the output of the crystal is given by
\begin{equation}
{c}_j (\mathbf{x}_i) = \int {\rm d} \mathbf{x}_j' h_j(
\mathbf{x}_j,\mathbf{x}_j') {b}_j (\mathbf{x}_j'), \label{cj}
\end{equation}
where ${b}_j (\mathbf{x}_j')$ are the Reference and Test
field operators at the output face of the crystal and
$h_R (\mathbf{x}_{R},\mathbf{x}_{R}')$ and $h_T
(\mathbf{x}_{T},\mathbf{x}_{T}')$ are the two
response functions describing the propagation of the field in
the two arms of the setup \cite{Goodman}.
\begin{widetext}
By using Eqs.~(\ref{cj}) and  (\ref{Corrfunc}) we can rewrite
$G^{(2)}(\mathbf{x}_R,\mathbf{x}_T)$ as
\begin{eqnarray}\label{G2hb}
G^{(2)}(\mathbf{x}_R,\mathbf{x}_T)&=& \int  {\rm d}  \mathbf{x}_R' ~
{\rm d} \mathbf{x}_R'' ~ {\rm d} \mathbf{x}_T' ~ {\rm d}
\mathbf{x}_T'' ~h_R (\mathbf{x}_{R },\mathbf{x}_{R}') ~ h_R^{*}
(\mathbf{x}_{R},\mathbf{x}_{R}'') ~ h_T
(\mathbf{x}_{T},\mathbf{x}_{T}')~  h_T^{*}
(\mathbf{x}_{T},\mathbf{x}_{T}'') \\ \nonumber & & \times
\left[\langle {b}_R^\dagger (\mathbf{x}_R'') {b}_R
(\mathbf{x}_R') {b}_T^\dagger (\mathbf{x}_T'') {b}_T
(\mathbf{x}_T') \rangle - \langle {b}_R^\dagger
(\mathbf{x}_R'') {b}_R (\mathbf{x}_R')\rangle \langle
{b}_T^\dagger (\mathbf{x}_T'') {b}_T (\mathbf{x}_T')
\rangle\right]
\end{eqnarray}
Also in this case the factorization rule for $\langle b_R^\dagger
(\mathbf{x}_R'') b_R (\mathbf{x}_R') b_T^\dagger (\mathbf{x}_T'')
b_T (\mathbf{x}_T') \rangle$ is the same as that for spontaneous PDC
\cite{Brambilla2004}, and for multi-thermal one-arm-seeded PDC
\cite{OPTLETT}
\begin{eqnarray}
\langle  {b}_R^\dagger (\mathbf{x}_R'') {b}_R
(\mathbf{x}_R') {b}_T^\dagger (\mathbf{x}_T'') {b}_T
(\mathbf{x}_T') \rangle = \langle {b}_R^\dagger
(\mathbf{x}_R'') {b}_R (\mathbf{x}_R')\rangle \langle
{b}_T^\dagger (\mathbf{x}_T'') {b}_T
(\mathbf{x}_T')\rangle +\langle {b}_R^\dagger (\mathbf{x}_R'')
{b}_T^\dagger (\mathbf{x}_T'')\rangle \langle {b}_R
(\mathbf{x}_R') {b}_T(\mathbf{x}_T') \rangle. \label{eqTH3}
\end{eqnarray}
The result in Eq.~(\ref{eqTH3}) can be demonstrated by rewriting ${b}_j (\mathbf{x})$ in terms
plane waves as ${b}_j (\mathbf{x}) \propto \sum_{\mathbf{q}}
e^{i \mathbf{q}\cdot \mathbf{x}}{b}_{j, \mathbf{q}}$ and
then exploiting the input-output relation of Eq. (\ref{aout}).
According to Eq. (\ref{eqTH3}), and in complete analogy with the case
of spontaneous PDC\cite{Brambilla2004}, also in the case of the
MMT-seeded PDC we obtain
\begin{equation}
G^{(2)}(\mathbf{x}_R, \mathbf{x}_T) = \left| \int {\rm d}
\mathbf{x}_R' \int {\rm d} \mathbf{x}_T'  h_R (\mathbf{x}_R,
\mathbf{x}_R') h_T (\mathbf{x}_T, \mathbf{x}_T') \langle
{b}_R
(\mathbf{x}_R') {b}_T(\mathbf{x}_T') \rangle \right|^2
\label{eqTH4}
\end{equation}
where
\begin{equation}
\langle {b}_R (\mathbf{x}_R') {b}_T(\mathbf{x}_T') \rangle
\propto  \sum_{\mathbf{q}} e^{i[\mathbf{q} \cdot
(\mathbf{x}_{T}'-\mathbf{x}_{R}')+\varphi_{\mathbf{q}}]}
~\mathcal{U}_{\mathbf{q}} \mathcal{V}_{\mathbf{q}} \left(1+
\mu_{T,\mathbf{q}}+ \mu_{R,-\mathbf{q}} \right) = \sum_{\mathbf{q}} e^{i[\mathbf{q} \cdot
(\mathbf{x}_{T}'-\mathbf{x}_{R}')+\varphi_{\mathbf{q}}]}
~\mathcal{C}_{\mathbf{q}} . \label{bb30}
\end{equation}
and $\mathcal{C}_{\mathbf{q}}$ is calculated in Eq.~(\ref{primati})
By using Eq.~(\ref{bb30}), Eq.~(\ref{eqTH4}) can be rewritten as
\begin{equation}
G^{(2)}(\mathbf{x}_R, \mathbf{x}_T) = \left|
\sum_{\mathbf{q}} \tilde{h}_R (\mathbf{x}_R,
-\mathbf{q}) \tilde{h}_T (\mathbf{x}_T,\mathbf{q})\mathcal{C}_{\mathbf{q}}
\right|^2
\label{eqTH4bis}
\end{equation}
where $\tilde{h}_j (\mathbf{x}_j,\mathbf{q}) =\int \mathrm{d}\mathbf{x}'_{j} e^{i\mathbf{q}\cdot \mathbf{x}'_{j}}
h_j (\mathbf{x}_j,\mathbf{x}'_{j})$.
\begin{figure}[h]
\includegraphics[width=0.45\textwidth,angle=270]{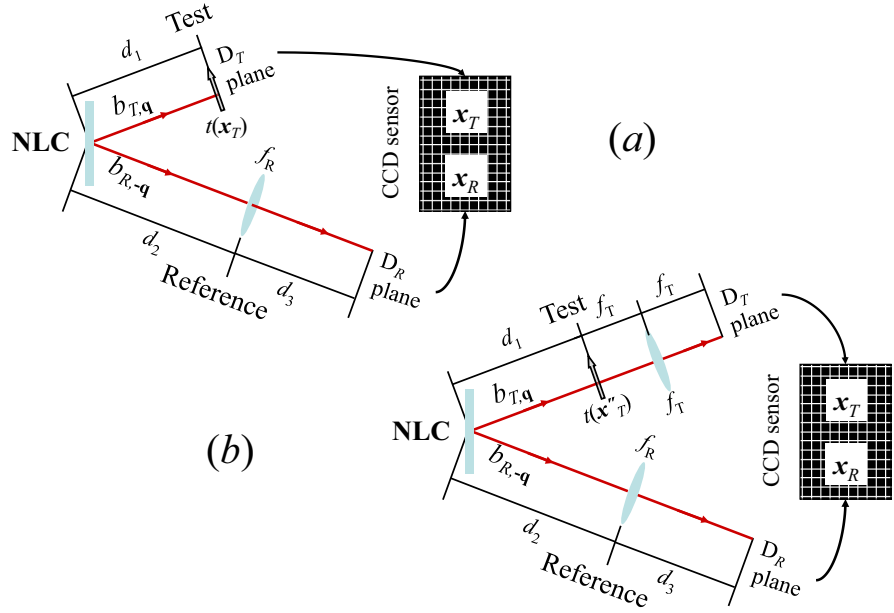}
\caption{Experimental setup for ghost imaging: $t(\mathbf{x_{T}})$, object
transmission function; $f_{T,R}$. (Left): experimental configuration with detection plane
coinciding with the object plane. (Right): experimental configuration with detection plane
coinciding with the Fourier plane of the collecting lens in the Test arm.}\label{setupGHOST}
\end{figure}
\par
According to Fig.~\ref{setupGHOST}, we  consider two different schemes for the  collection
optics in the Test arm of the setup:
\par\noindent
($a$) the detection plane coincides with the plane of the
transparency, $\mathbf{x}''_{T}=\mathbf{x}_{T}$, and hence
\begin{eqnarray}
\tilde{h}_T (\mathbf{x}_{T},\mathbf{q})\propto e^{-i\frac{\lambda d_1}{4\pi} q^2} e^{-i\mathbf{q}\mathbf{x}_{T}}  t(\mathbf{x}_{T})
\label{eqhT1}
\end{eqnarray}
only describes free propagation over a distance $d_{1}$;
\par\noindent
($b$) a collection lens is located on the plane $\mathbf{x}_{l,T}$ beyond the transparency in a Fourier-transform configuration, and hence
\begin{eqnarray}
\tilde{h}_T(\mathbf{x}_{T},\mathbf{q})\propto e^{-i\frac{\lambda d_1}{4\pi} q^2} \tilde{t}(-\mathbf{q}-\frac{ 2\pi}{ \lambda f_T}\mathbf{x}_{T} ) \label{eqhT2}\; .
\end{eqnarray}
The optical scheme in the Reference arm contains a lens (focal length $f_R$) located at  $\mathbf{x}_{l,R}$ and thus the Fourier transform of the impulse response functions can be written as
\begin{eqnarray}
\tilde{h}_R(\mathbf{x}_{R},-\mathbf{q}) &\propto&
\int {\rm d} \mathbf{x}'_R e^{-i \mathbf{q}  \cdot \mathbf{x}'_R}
\int {\rm d} \mathbf{x}_{l,R}
e^{i\frac{\pi }{\lambda d_2}
(\mathbf{x}_{l,R}-\mathbf{x}'_{R} )^{2}}
e^{-i\frac{\pi }{\lambda f_R}\mathbf{x}^{2}_{l,R}}e^{i\frac{\pi }{\lambda d_3}
(\mathbf{x}_{l,R}-\mathbf{x}_{R})^{2}}\nonumber\\
&\propto& e^{-i\frac{\lambda d_2}{4\pi} q^2}\int {\rm d} \mathbf{x}_{l,R}
e^{-i\left(\frac{2\pi }{\lambda d_3}\mathbf{x}_{R}+\mathbf{q}\right) \cdot\mathbf{x}_{l,R}}
e^{i\frac{\pi}{\lambda}\left(\frac{1}{d_3}-\frac{1}{f_R} \right)\mathbf{x}_{l,R}^{2}}. \label{eqhhR}
\end{eqnarray}
If $f_R\neq d_3$, Eq.~(\ref{eqhhR}) becomes
\begin{eqnarray}
\tilde{h}_R(\mathbf{x}_{R},-\mathbf{q}) &\propto&
e^{-i\frac{\lambda}{4\pi}\left(d_2+\frac{1}{1/d_3-1/f_R} \right){q}^{2}}
e^{-\frac{i}{d_3}\frac{1}{1/d_3-1/f_R}\mathbf{q} \cdot \mathbf{x}_{R}}\; , \label{eqhhRim}
\end{eqnarray}
while if $f_R = d_3$, Eq.~(\ref{eqhhR}) becomes
\begin{eqnarray}
\tilde{h}_R(\mathbf{x}_{R},-\mathbf{q}) &\propto& e^{-i\frac{\lambda d_2}{4\pi} q^2} \delta\left(\frac{2\pi }{\lambda d_3}\mathbf{x}_{R}+\mathbf{q}\right)\; . \label{eqhhRfour}
\end{eqnarray}
Depending on the chosen geometrical configuration, these schemes
realize either a ghost-imaging or a ghost-diffraction experiment
\cite{GIGD}.
\subsection{Ghost imaging} To perform a ghost-imaging experiment we
choose $f_R\neq d_3$. Let us first consider case ($a$). Substituting
Eq.~(\ref{eqhT1}) and Eq.~(\ref{eqhhRim}) into Eq.~(\ref{eqTH4bis})
yields the expression
\begin{eqnarray}
G^{(2)}(\mathbf{x}_R, \mathbf{x}_T) &\propto& \left| t(\mathbf{x}_{T})\right|^2\left|
\sum_{\mathbf{q}} \mathcal{C}_{\mathbf{q}} e^{-i\mathbf{q}\cdot \left( \mathbf{x}_{T}+\frac{1}{d_3}\frac{1}{1/d_3-1/f_R}\mathbf{x}_{R}\right)}
e^{-i\frac{\lambda}{2\pi}\frac{d_1+d_2}{1/d_3-1/f_R}\left(\frac{1}{d_1+d_2}+ \frac{1}{d_3} - \frac{1}{f_R} \right){q}^{2}}
\right|^2 \nonumber\\
&\simeq& \left| t(\mathbf{x}_{T})\right|^2\left|\mathcal{C}_{\mathbf{q}}\right|^2\delta\left(\mathbf{x}_{T}+\frac{\mathbf{x}_{R}}{M}\right)
\label{eqTH4bis2}\; ,
\end{eqnarray}
which, once integrated over the bucket detector,
\begin{eqnarray}
G^{(2)}(\mathbf{x}_R)=\int {\rm d} \mathbf{x}_T  G^{(2)}(\mathbf{x}_R, \mathbf{x}_T) \simeq  \left| t(-\frac{\mathbf{x}_{R}}{M})\right|^2\left|\mathcal{C}_{\mathbf{q}}\right|^2
\label{eqTH4ter}\; ,
\end{eqnarray}
gives the image of the object. Note that in passing from Eq.~(\ref{eqTH4bis}) to Eq.~(\ref{eqTH4ter}) we
have made the following assumptions: $\mathcal{C}_{\mathbf{q}}$ is almost independent of ${\mathbf{q}}$ and
the distances $d_1$, $d_2$ and $d_3$ satisfy the so-called ``back-propagating thin lens
equation'', $1/(d_1+ d_2)+1/d_3=1/f_R$ \cite{Abouraddy2002}, so that we obtain an imaging system with
magnification factor $M = d_3/(d_1+d_2)$.
In case ($b$), that is with a collection lens in the Test arm we proceed similarly by substituting
Eq.~(\ref{eqhT2}) and Eq.~(\ref{eqhhRim}) into Eq.~(\ref{eqTH4bis}) and making the same assumptions as before:
\begin{eqnarray}
G^{(2)}(\mathbf{x}_R, \mathbf{x}_T) &\propto& \left|
\sum_{\mathbf{q}} \mathcal{C}_{\mathbf{q}} \tilde{t}(-\mathbf{q}-\frac{ 2\pi}{ \lambda f_T} )e^{-i\frac{1}{d_3}\frac{1}{1/d_3-1/f_R}\mathbf{q}\cdot \mathbf{x}_{R}}
e^{-i\frac{\lambda}{2\pi}\frac{d_1+d_2}{1/d_3-1/f_R}\left(\frac{1}{d_1+d_2}+ \frac{1}{d_3} - \frac{1}{f_R} \right){q}^{2}}
\right|^2\nonumber\\
&\simeq& \left| t(-\frac{\mathbf{x}_{R}}{M})\right|^2\left|\mathcal{C}_{\mathbf{q}}\right|^2
\; .
\label{eqTH4quater}
\end{eqnarray}
Note that in this case ($b$) the image of the object emerges from correlations without need to perform the integration over the bucket detector in order to recover the ghost image.
\subsection{Ghost diffraction} To perform a ghost-diffraction
experiment we consider the configuration $d_3=f_R$ and choose
configuration ($b$) in the Test arm. By inserting Eq.~(\ref{eqhT2})
and Eq.~(\ref{eqhhRfour}) into Eq.~(\ref{eqTH4bis})
\begin{eqnarray}
G^{(2)}(\mathbf{x}_R, \mathbf{x}_T) &\propto& \left|
\sum_{\mathbf{q}} \mathcal{C}_{\mathbf{q}} \tilde{t}(-\mathbf{q}-\frac{ 2\pi}{ \lambda f_T} )
e^{-i\frac{\lambda}{4\pi}\left(d_1+d_2\right){q}^{2}}\delta\left(\frac{2\pi}{\lambda d_3}\mathbf{x}_R-\mathbf{q}\right)
\right|^2\nonumber\\
&\simeq& \left| \tilde{t}(-\frac{2\pi}{\lambda d_3}\mathbf{x}_{R}-\frac{ 2\pi}{ \lambda f_T}\mathbf{x}_{T})\right|^2\left|\mathcal{C}_{\mathbf{q}}\right|^2
\; .
\label{eqTH4diffr}
\end{eqnarray}
By selecting the component $\mathbf{x}_T=0$ on the Test plane we get
\begin{eqnarray}
G^{(2)}(\mathbf{x}_R, 0)
&\simeq& \left| \tilde{t}(-\frac{2\pi}{\lambda d_3}\mathbf{x}_{R})\right|^2\left|\mathcal{C}_{\mathbf{q}}\right|^2
\; ,
\end{eqnarray}
which gives the diffraction pattern of the object. Note that the choice ($a$) would not give any meaningful result.
\end{widetext}
\section{Conclusions and outlooks}
\label{s:out}
This paper was aimed at showing the possibility of performing ghost
imaging and ghost diffraction with a novel source based on PDC seeded
with two MMT fields, which generates a bipartite correlated state.
Peculiar properties of this new source may open a new insight into the
understanding of the ghost imaging/diffraction process. In fact,
nowadays the sources considered for ghost imaging/diffraction either
were definitely separable (classically correlated beams obtained from a
MMT source) or entangled (spontaneous PDC). On the contrary, here we
proved that the separable/entangled nature of the light produced by our
source can be controlled by changing the seed intensities and that the
transition from quantum to classical regimes does not modify the
possibility of realizing ghost imaging schemes.
\par
Furthermore, we also showed that a ghost imaging experiment performed
with our source satisfies the``back-propagating'' thin-lens equation, as
much as with spontaneous PDC, even when the state produced becomes
separable. This is in contrast with the idea, also recently suggested
\cite{Dez05,Cai05,Cai05_2}, that the ``back-propagating'' thin-lens
equation is connected with the entangled nature of the spontaneous PDC.
\par
According to the consideration of above, we are planning to realize
a ghost imaging experiment with a MMT seeded PDC source in order to
show that the same optical configuration allows retrieving of the
image irrespectively to the entangle or separable nature of the
light produced by the source. This will definitely demonstrate that
there is not any connection between the entangled properties of the
light source and the ``back-propagating'' thin-lens equation.
\section*{Acknowledgments}
This work has been supported by MIUR project PRIN2005024254-002.

%
\end{document}